# High-Efficiency Multilevel Phase Lenses with Nanostructures on Polyimide Membranes

Leslie Howe,[1] Tharindu D. Rajapaksha,[1] Kalani H. Ellepola,[1] Vinh X. Ho,[1] Zachary Aycock,[1] Minh L. P. Nguyen,[1] John P. Leckey,[2] Dave G. Macdonnell,[2] Hyun Jung Kim,[2] and Nguyen Q. Vinh[1]*

[1] Department of Physics and Center for Soft Matter and Biological Physics, Virginia Tech, Blacksburg, VA 24061, USA

[2] NASA Langley Research Center, Hampton, Virginia 23681, USA

* Corresponding author: vinh@vt.edu; phone: 1-540-231-3158

**ABSTRACT**

The emergence of planar meta-lenses on flexible materials has profoundly impacted the long-standing perception of diffractive optics. Despite their advantages, these lenses still face challenges in design and fabrication to obtain high focusing efficiency and resolving power. A nanofabrication technique is demonstrated based on photolithography and polyimide casting for realizing membrane-based multilevel phase-type Fresnel zone plates (FZPs) with high focusing efficiency. By employing advantageous techniques, these lenses with nanostructures are directly patterned into thin polyimide membranes. The computational and experimental results have indicated that the focusing efficiency of these nanostructures at the primary focus increases significantly with increasing the number of phase levels. Specifically, 16-level phase lenses on a polyimide membrane can achieve a focusing efficiency of more than 91.6% of the input signal (9.5 times better than that of a conventional amplitude-type FZP) and focus light into a diffraction-limited spot together with very weak side-lobes. Furthermore, these lenses exhibit considerably reduced unwanted diffraction orders and produce extremely low background signals. The potential impact of these lenses extends across various applications and techniques including microscopy, imaging, micro-diffraction, remote sensing, and space flight instruments which require lightweight and flexible configurations.

# 1. Introduction

Planar diffractive optical elements (DOEs) exhibit excellent imaging and focusing capabilities across the entire electromagnetic spectrum from X-rays to visible, infrared, and terahertz.[1-5] Compared to reflective and refractive optical components, planar DOEs have many advantages such as small volume, lightweight, special dispersion characteristics, sub-diffraction focal spot size, and greater freedom in optical design for specific applications,[3, 4, 6-10] These optics can be employed in a wide range of applications including X-ray focusing,[11] microscopy,[6, 11] spectroscopy,[12] nanolithography,[13] optical communication devices,[14] microwave antennas,[15] beam shaping,[16-18] optical interconnects,[14] imaging,[19] remote sensing,[20] and next-generation space telescopes and instrumentation.[8] Flight instruments, including specifically light detection and ranging (LiDAR) systems, will benefit from increasingly large apertures.[8, 21] Reflective and refractive instruments will be impractical in such applications. Large aperture DOEs are lightweight, replicable, and deployable, which makes them desirable candidates for such optical instruments.[8, 22-24] However, the fabrication and optical performance of DOEs such as focusing efficiency and resolving power, remain complex.[2-4, 25]

Diffraction occurs in all phenomena involving electromagnetic waves, although it is not usually dominant; thus, produces a weak contribution and establishes physical limitations.[1, 26, 27] As an example, for ideal optical systems without any aberrations or imperfections, the smallest focal spot size of a refractive lens is described as the diffraction limit, theoretically limited by the Rayleigh criterion of $0.61\times\lambda/NA$ (where NA is numerical aperture and $\lambda$ is the wavelength).[1] DOEs can achieve optical imaging resolution below the diffraction limit.[28, 29] However, unlike refractive lenses where incident radiation is bent into one location, DOEs generate multiple diffraction orders at different positions and side-lobes, significantly reducing the focusing efficiency at the main focus. By modifying the structure of DOEs including multilevel diffractive structures,[4, 11, 28-32] applying metamaterial artificial structures including active, electrically-tunable metasurfaces,[33-35] or designing metasurfaces including plasmonic and dielectric meta-lenses,[36-39] we can obtain a high focusing efficiency, increase the resolving power or control tunable metasurfaces.[4, 11, 28-44] Specifically, DOEs can obtain a sub-diffraction focal spot such as supercritical[28, 29], superoscillatory lenses,[40-43] and hyperlenses[44] but at the cost of reduced focusing efficiency and elevated intensity in side-lobes. Therefore, depending on the application, we can optimize the performance of DOEs through the modification of their structures. One of the most challenging aspects of DOEs is to obtain high focusing efficiency,[45-48] because a considerable fraction of incident beam is focused into non-desirable orders, consequently significantly reducing the signal-to-noise ratio at the plane of interest.[1, 27] With a careful design, DOEs can be engineered to obtain a desired intensity and phase profile at the focal plane.[11, 23, 46, 49] The focusing efficiency of phase-type DOEs can theoretically reach 100% with a continuous surface-relief profile at a desirable plane.[27, 50] Specifically, destructive interference in the zeroth and negative diffraction orders can be obtained in phase-type DOEs. Theoretical estimation of the focusing efficiency of phase-type FZPs with a multilevel profile or multilevel FZPs (MLFZPs) reveals significant improvements as the number of phase levels increases.[8, 11, 51] For example, when the phase level increases from 2 to 4 levels, the theoretical focusing efficiency of a MLFZP will grow from 40.5 to 81.1%.[51] Thus, understanding diffraction and demonstrating diffractive optics are crucial for advances in various scientific and engineering fields.

Thin optical membranes are a solution for making large, launchable, flexible, lightweight, compactly packageable and deployable optical instruments. However, it is difficult to control vertical sidewall shapes and maintain the accuracy of zone-profiles during the nanofabrication of DOEs.[8, 22, 52] Accordingly, the focusing efficiency of the focal spot is reduced and the intensity profile becomes broader, along with an increase in the intensity in other diffraction orders from imperfections in the fabrication of DOEs. Several methods have been employed for DOE fabrication including single-point diamond turning, electron-beam (e-beam) lithography, photolithography combined with dry etching, and laser pattern generation to produce various diffraction features.[11, 23, 46, 53] E-beam lithography provides exceptionally high-resolution sub-micrometer features but requires expensive and complex equipment, and an extremely long exposure time to cover a large area. On the other hand, the precise transfer of nanostructures onto substrate materials remains a significant challenge in laser-direct writing processes. Additionally, uniformity of photoresist thickness is difficult to realize, especially on a large-aperture substrate.[50, 54]

In this work, we have demonstrated a significant improvement in the focusing efficiency of polyimide membrane MLFZPs fabricated by employing nanofabrication techniques that combine photolithography and polyimide casting. These MLFZP lenses obtain a high focusing efficiency into a diffraction-limit spot, significantly reduced intensity at side-lobes and unwanted diffraction orders, thus, having low background signal. Theoretical calculations and experimental results have illustrated that the polyimide membrane MLFZPs can achieve a focusing intensity of more than 91.6% of the input signal, a significant improvement when compared with that of a conventional binary amplitude-type FZP. With these unique properties, the DOEs can make a significant impact on research and optical instrument development including microscopy, micro-diffraction, LiDAR, imaging, spectroscopy, instruments, spaceborne astrophysics, and planetary science.

## 2. Nanofabrication and optical characterization

One common use of diffractive lenses such as FZPs is to bring light from a distant source to a focus. Light waves passing through transparent rings of a FZP diffract at the edges of the rings, generating spherical wavefronts as the Huygens-Fresnel wave propagation principle states, and then combine at the focus.[1, 26] The total contribution of the field at the focus is constructed by summing fields from all points on the wavefront with their respective phases to the focal point, determined only by optical path lengths between the point positions on the diffractive element and the focus. If the optical path lengths of two points on the wavefronts to the focus differ by an even number of half-wavelengths, the two-point sources will interfere constructively (in phase); inversely, if the optical path length difference is an odd number of half-wavelengths, they will interfere destructively (out of phase). Thus, adjacent rings will transmit waves with opposite signs of phase. If alternate rings are blocked, then the contribution at the focus will have the same phase sign. The total intensity at the focus is proportional to the number of transparent rings, forming a diffractive amplitude-type FZP device. Two adjacent rings including a transparent and opaque area form a complete zone. The radius, $R_n$, of the $n^{th}$ transparent ring ($n$ may have only even values) of a FZP with focal length, $f$, and design wavelength, $\lambda$, is given by:[1, 27, 55]

$$R_n = \sqrt{nf\lambda + (n\lambda/2)^2} \qquad (1)$$

The focusing efficiency of an amplitude-type FZP is low. A FZP with equal areas of transparent and opaque rings has half of the incident intensity blocked by the opaque areas and only a small fraction of the incident energy is diffracted to the primary focus.[51] Because of their poor focusing efficiency, amplitude-type DOEs are not commonly used in commercial systems.

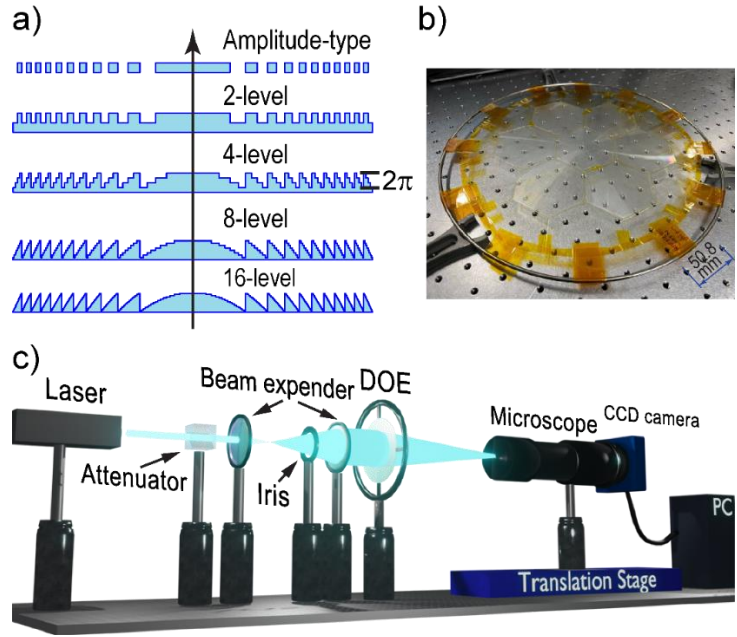

**Figure 1:** Schematic diagrams of different diffractive optic structures and the optical characterization setup. a) DOE patterns include amplitude-type and phase-type FZPs containing 2-, 4-, 8, and 16-level structures. b) MLFZPs are fabricated from casting polyimide membrane on silicon molds. c) The optical setup for optical characterization of these MLFZPs includes a beam expander, custom microscope, and CCD camera.

To obtain a high focusing efficiency from diffractive elements, a phase-type FZP is realized in which opaque rings in the amplitude-type FZP are replaced by phase-shift regions. Specifically, the thickness of the phase-shift region is designed to create a half-wavelength phase shift (or a phase change of π), which provides a proper phase for constructive interference at the focal point. However, a ring has a physical width, and as a result, the phase of light at the focus passing near the inner edge of the ring is different from that of light passing near the outer edge of the ring. Therefore, a MLFZP is constructed by dividing a zone or pair of transparent and opaque rings into a gradient of phase shifts. A phase shift of 2π in one zone can be obtained by varying the thickness of the structure. Schematics of phase-type DOEs for 2-, 4-, 8- and 16-levels are illustrated in Figure 1a. The thickness of each phase level, *d*, is given by:[11, 46]

$$d = \frac{\lambda}{N\,(n_{\text{sub}} - n_{\text{env}})}, \qquad (2)$$

where *N* is the number of phase levels, $n_{\text{env}}$ and $n_{\text{sub}}$ are refractive indices of the surrounding environment and substrate, respectively. The zone radius of the $k^{\text{th}}$ level in the *N*-level phase structure within the $(n/2)^{\text{th}}$ zone is given by,[11]

$$R(k,n) = \sqrt{f\lambda\left(\frac{2k}{N} + n - 2\right) + \left(\frac{2k}{N} + n - 2\right)^2 \left(\frac{\lambda}{2}\right)^2}, \qquad (3)$$

where *k* is the sub-zone index ranging from 1 to *N*.

The focusing efficiency can be simply estimated based on multilevel phase-type gratings at the primary focus:[51, 56]

$$\eta = \frac{\sin^2(\pi/N)}{(\pi/N)^2}. \qquad (4)$$

For 2-, 4-, 8-, and 16-level phase-type structures, theoretical values of the focusing efficiency at the 1st diffraction order are 40.5, 81.1, 95.0 and 98.7%, respectively. Details about the focusing efficiency for different diffraction modes of polyimide membrane MLFZPs have been calculated by Fourier analyses.[27]

We have designed DOEs with a focal length of 3 m and a diameter of 0.080 m. Specifically, the amplitude-type FZP and phase-type MLFZPs with different diffraction levels (*N* = 2, 4, 8, 16), containing 491 concentric zones were constructed. The amplitude-type FZP was fabricated on quartz, containing metal concentric (opaque) rings to produce constructive interference at the various foci.[57-59] The polyimide membrane phase-type MLFZPs with different nanostructures have been fabricated on high-performance CORIN® XLS Polyimide material.[60] Figure 1b shows a polyimide membrane sheet containing seven hexagons with different nanostructures on MLFZPs (one 16-level FZP at the center and six blank outside hexagons). Arranging components in a hexagon configuration can enable fabricating a large aperture beyond the capability of a single nanofabrication instrument. The fabrication has been demonstrated using 4-inch wafer instruments. Designing and building optical components larger than the limitation of the instrument can be obtained using the hexagon approach. To compare optical performance, the DOEs are fabricated with the same diameter and focal length (or $f_{number}$). The total thickness of the polyimide membrane is approximately 30 μm. The material is a clear, colorless organic/inorganic nanocomposite with exceptional resistance to atomic oxygen erosion and the refractive index of 1.54 ± 0.01 at 532 nm. The polyimide exhibits high heat resistance with a glass transition temperature of more than 266 °C, making it suitable for many high temperature applications.

The DOEs were fabricated using photolithography and polyimide membrane casting. The amplitude-type FZP was fabricated directly using a photolithography process with a thin chromium metal layer of 75 nm for opaque rings on a 100-mm diameter and 500-μm thick quartz wafer (Figure S1). The phase-type MLFZPs were fabricated by casting CORIN polyimide on silicon molds. Figure 2 shows the fabrication steps for a MLFZP silicon mold and the subsequent polyimide casting. To fabricate silicon molds, advanced photolithography techniques are employed on 100-mm diameter and 500-μm thick silicon wafers and followed by dry reactive-ion etching techniques. The depth of each level on the silicon molds for different MLFZPs was initially estimated from Eq. 2 with the refractive index of CORIN material; however, the membrane thickness shrinks after the casting process at high temperature. We optimized all the processes and obtained the optimal depth, $d$, for silicon molds ≈ 3% deeper than those from theoretical estimations. A 2-level FZP photomask was used to make 2-level silicon molds, after that a 4-level FZP photomask was employed to etch the previous silicon mold into a 4-level structure. The process was continued for 8- and 16-level silicon molds (see Supporting Information). The polyimide casting was performed at Nexolve using a proprietary process including dissolving the polyimide, spin coating the solution into the molds, heat curing, releasing, and post-processing the membranes.[60]

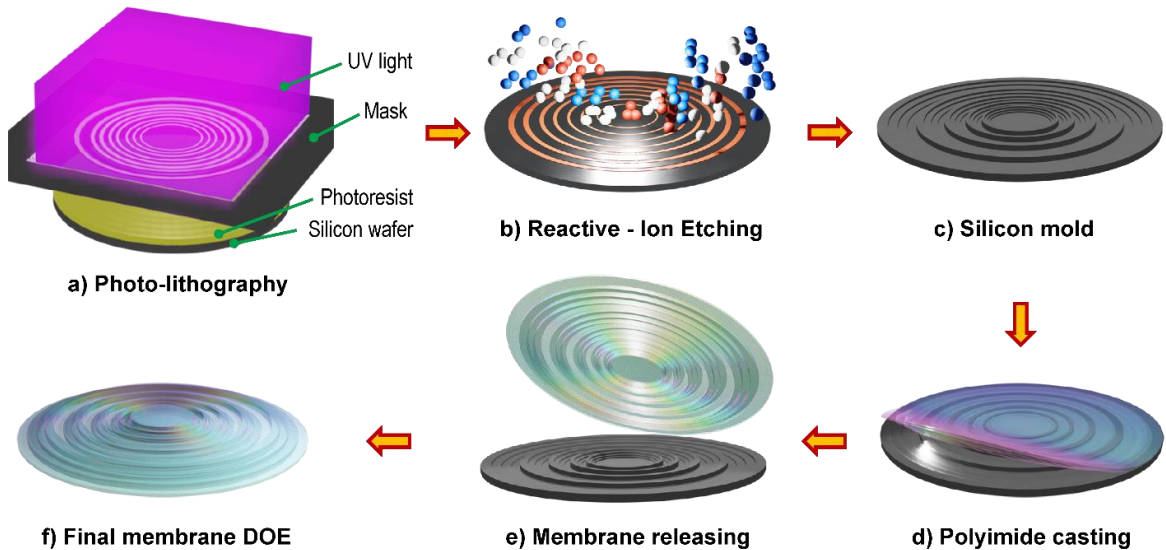

**Figure 2:** Fabrication procedure for DOEs includes a) photolithography, b) reactive ion etching, c) silicon mold, d) polyimide membrane casting, e) membrane releasing, and f) final membrane DOE.

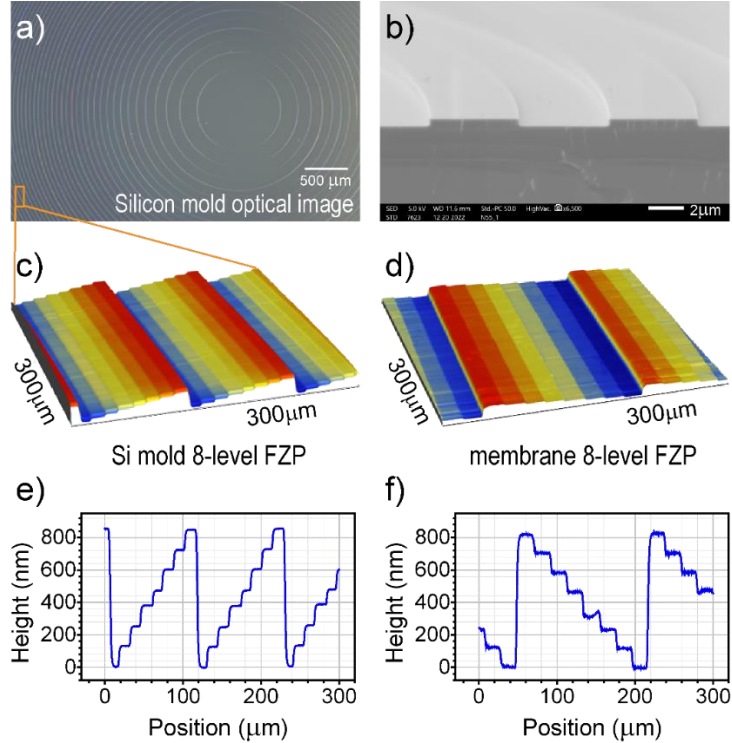

**Figure 3:** Physical characterization of DOEs. a) Optical image of silicon mold provides a quick view of DOEs. b) SEM image can evaluate nanostructures of DOEs. c,d) AFM images of a silicon mold and polyimide membrane 8-level FZP, respectively, show three-dimensional nanostructures of DOEs. e,f) vertical profiles of the silicon mold and polyimide membrane 8-level FZP, respectively, illustrate nano-steps on these DOEs.

We have performed physical characterizations of the silicon molds and cast polyimide membrane MLFZPs to understand their impact on the optical performance of these DOEs. For a quick evaluation of these structures, an optical microscope can provide images of the surface structure on these optical elements (Figure 3a). To confirm the height of these nanostructures, scanning electron microscopy (SEM) was used (Figure 3b). Furthermore, atomic force microscope (AFM) images combined with vertical profiles of these structures are shown in Figure 3, illustrating the nanofeatures on these DOEs. AFM images of silicon molds provide the processing error for the non-uniformity of the lithography process of between 0.5 and 3.0%, depending on the wafer and inherent physics in the dry-etching process, in agreement with previous reports using reactive ion etching (RIE).[58, 61] The uniformity of AFM images of MLFZPs shows a similar result, demonstrating a highly accurate casting process. In addition, the SEM images indicate an etch anisotropy of 0.95 ± 0.03, which is close to 1, under our process used here (≈4 nm/s etch rate together with other conditions presented in Supporting Information), revealing an accurate combination of the silicon lithography/nanofabrication and polyimide casting processes. We consider this minimal deviation from perfection to be proof of minimal processing errors. The focusing efficiency observed of the optical elements and their deviation from the theoretically calculated values is the optical characterization resulting from the minimal manufacturing discrepancies.

To evaluate the optical performance of the DOEs, we have built an optical setup to characterize focusing and imaging properties. A schematic diagram of components used to obtain experimental results is shown in Figure 1c. A 532-nm laser beam produced by a Coherent Verdi – 2 W laser system is steered through mirrors to a beam expander which includes two lenses with their focal lengths of 3.9 and 477 mm. The beam expander produces a homogenous beam with a diameter of ≈ 200 mm. A large beam diameter ensures that the beam incident on a DOE under test does not contain a significant amount of diffracted light. An attenuator is placed in front of the beam expander to adjust the intensity to avoid over-saturation of the detector. An iris between the two lenses has been used to control the diameter of the beam before impacting a DOE. In addition, the DOEs were placed at different positions from the sources during the testing, which would produce differences in phase of the light source incident to the lenses and resulted in no change in efficiency. The image at the focal plane of the DOE has a small spot size; therefore, a custom microscope was built to magnify these images for proper characterization. The diffracted light from the DOE is incident on the microscope with different magnifications from 10 × to 80 × and is digitized using CMOS cameras (monochrome BFS-U3-200S6M-C with pixel size of 2.4 × 2.4 µm and color DCC1645C with pixel size of 3.6 × 3.6 µm). With the help of the microscope, small features of focused light at the focal plane can be observed. The microscope was placed on a long translation stage to capture images along the optical axis. To measure the focusing efficiency, the microscope was then replaced by a power meter. A proper pinhole was used to obtain the intensity of focusing light within a spot of diameter equal to 3 times the simulated full-width at half-maximum (FWHM).[4] The focusing efficiency is determined as the ratio of the intensities of the diffracted light at the focus to the incident light on the DOE.

## 3. Numerical simulations

Optical elements, interfaces, and surfaces change light properties including direction, phase, amplitude, and polarization through optical processes of reflection, refraction, absorption, and diffraction. The magnitude of the electric field of light at the focal plane can be determined using the Rayleigh–Sommerfeld diffraction theory.[62] For spherical wavefronts, the scalar diffraction formula for the electric field at point P on an observation ($x$, $y$) plane, which is contributed from the whole aperture area, $\Sigma$, is given by the Fresnel – Kirchoff diffraction formula,[55, 63]

$$E(P) = \frac{1}{i\lambda} \iint_\Sigma E_0(\xi, \eta, 0) \frac{e^{iks}}{s} \left[\frac{1+\cos\theta}{2}\right] d\xi d\eta , \qquad (5)$$

where the factor in square brackets, $(1 + \cos\theta)/2$, is known as the Kirchhoff obliquity factor. This factor is reduced to $\cos\theta$ for the first Rayleigh – Sommerfeld solution and approximated to 1 for the second Rayleigh – Sommerfeld solution. The Huygens – Fresnel principle, as predicted by the first Rayleigh – Sommerfeld solution is known as the Fresnel – Kirchhoff integral,[27, 55, 63]

$$E(P) = \frac{1}{i\lambda} \iint_\Sigma E_0(\xi,\eta,0) \frac{e^{iks}}{s} \cos\theta \, d\xi d\eta = \frac{z}{i\lambda} \iint_\Sigma E_0(\xi,\eta,0) \frac{e^{iks}}{s^2} \, d\xi d\eta. \qquad (6)$$

The Fresnel - Kirchhoff integral is difficult to evaluate analytically, even when the obliquity factor is approximated as one (i.e. far forward propagation) and a uniform field is employed at the aperture (i.e. constant). For small angle scattering in the near-forward direction, Fresnel approximated $s \cong z$ in the denominator. The radius of each wavelet approximated as, $s = \sqrt{z^2 + (x-\xi)^2 + (y-\eta)^2} \simeq z\left[1 + \frac{1}{2}\left(\frac{x-\xi}{z}\right)^2 + \frac{1}{2}\left(\frac{y-\eta}{z}\right)^2\right]$. For $z \gg x, y$ and $z \gg \xi, \eta$, the Fresnel approximation is derived,

$$E(P) = \frac{e^{ikz}}{i\lambda z} \iint_\Sigma E_0(\xi,\eta,0) \, e^{\frac{ik}{2z}[(x-\xi)^2+(y-\eta)^2]} d\xi d\eta. \qquad (7)$$

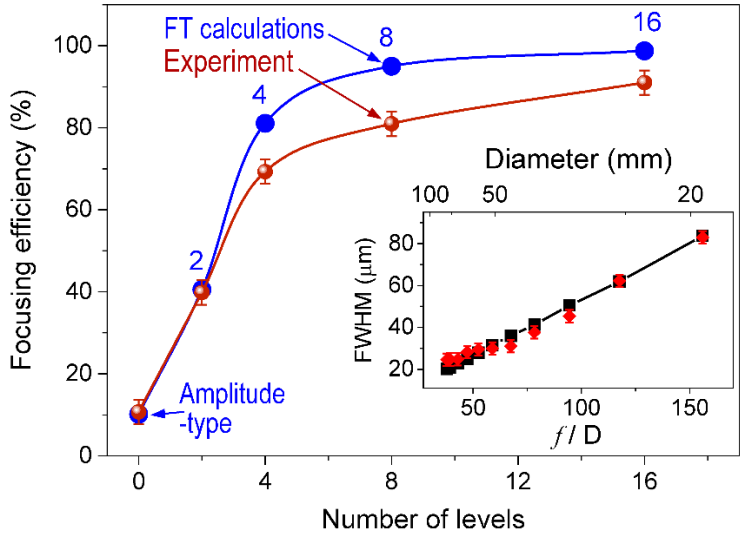

**Figure 4:** Focusing efficiency of MLFZPs. The focusing efficiency increases with the number of phase levels. The highest focusing efficiency ($\approx 91.6 \pm 3.0\%$) has been obtained for the 16-level FZP, which is 9.5 times higher than that of the conventional amplitude-type FZP. (Inset) the focal spot size (FWHM) decreases with reducing the $f_{number}$. The black line is the FWHM extracted from FFT calculations and red symbols are the optical characterizations.

We have performed numerical simulations for DOEs using the fast-Fourier transform (FFT) in MATLAB.[27, 64, 65] Specifically, the DOEs are defined as an ensemble of concentric rings with different phases and amplitude (Figure 1a). Complex values ($Ae^{i\Phi}$) are assigned to the matrix of pixels of the DOEs, where *A* is the amplitude transmittance, and $\Phi$ is the phase value. In the case of the amplitude-type DOEs, the transparent and opaque rings form areas with the transmittance of '1' and '0', respectively, of the photon flux, and the phase, $\Phi$, is a constant or zero. Whereas in the case of phase-type DOEs, the entire area has a constant transmittance, *A*, and the phase is divided into *N* levels. The diffraction pattern of the DOE on vertical planes along the optical axis (*z*) is calculated numerically by FFT using the scalar Fresnel diffraction approximation. The calculations have been carried out for a large number of zones (491 in total), which form DOE lenses with a diameter of 80 mm. We also varied the diameter of the DOEs to explore the diffraction limit of the focal spot size of these optical elements.

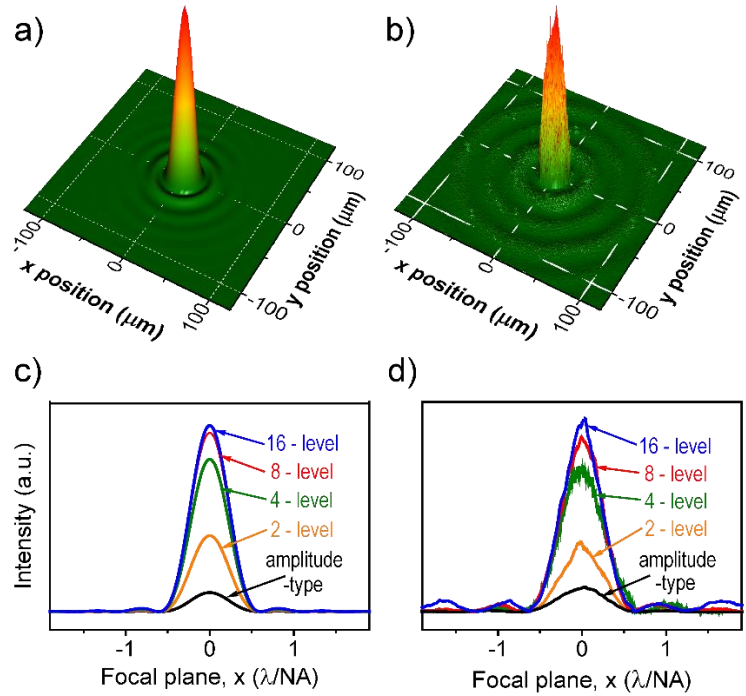

**Figure 5:** Intensity profiles at the focal planes of 16-level MLFZPs. a) Diffraction fringes computed using the Huygen – Fresnel approximation. b) Measured diffraction fringes at the primary focal plane, c,d) Comparison of intensity profiles of MLFZPs with different diffraction structures, including the amplitude-type, multilevel phase-type FZPs.

## 4. Results and discussion

The greatest advantages of our fabrication technique are the ability to obtain a high focusing efficiency of phase-type MLFZPs on polyimide membranes and the ability to fabricate large optical elements (Figure 1), satisfying the requirement of constructing flexible and lightweight optical elements. To demonstrate the improved focusing efficiency, we have performed numerical calculations and characterized the optical properties of these nanostructures. The intensity reaches a maximum at the primary focus of these DOEs at 3 m for FFT calculations and experimental results, as the DOEs are specifically designed to operate at 532 nm with a 3-m focal length. Our experimental results have demonstrated that while the amplitude-type DOE has a focusing efficiency of 9.7%, the polyimide membrane MLFZPs have significantly increased focusing efficiencies at the primary focus (Figure 4). Specifically, focusing efficiencies of 38.7 ± 3.0, 67.9 ± 3.0, 80.8 ± 3.0 and 91.6 ± 3.0% have been obtained at the primary focus for 2-, 4-, 8-, and 16-level phase-type structures, respectively. These values are close to the theoretical calculations of ideal MLFZP structures, indicating an accurate fabrication process. The focusing efficiencies of ideal MLFZPs can be simply estimated using Eq. 4 for different phase-type grating structures. In addition, by employing the FFT in MATLAB and complex values of amplitude and phase of the DOEs, focusing efficiencies at the primary focus of 10.1, 40.5, 81.1, 95.0 and 98.7% (Figure 4) for the amplitude-type, 2-, 4-, 8-, and 16-level phase-type structures, respectively, were obtained.

The focusing efficiency of phase-type DOEs increases significantly with increasing the number of phase levels, which is realized from both experimental results and theoretical calculations. Figure 5 shows the intensity profile at the primary focal plane (FFT calculations and experiment results for the 16-level FZP) by employing the previously described setup of a collimated 532-nm laser beam with an 80-mm diameter on the center of the structure, corresponding to a DOE with 491 zones. The angular distribution of intensity on the primary focal plane at a distance, $z$, has a Bessel function in the form,[1, 55]

$$I(\theta) \propto E(P) \times E^*(P) \propto |2J_1(kR_n\theta)/(kR_n\theta)|^2 \qquad (8)$$

where $\theta \approx r/z$ is the angle from the polar axis at a distance, $r$, from the center point, $E(P)$ is the electrical field strength at the point P in the observation screen, * denotes its complex conjugate, $J_1(kR_n\theta)$ is the first kind, order one Bessel function and $k = 2\pi/\lambda$. An excellent agreement between simulations and experimental results has been achieved.

The main difference in phase-type MLFZPs is the peak intensity values due to the total amount of light converging at the primary focus (Figure 5 c,d). Similar to previous results obtained from focusing efficiency measurements (Figure 4), the peak measurement intensity taken from a CCD image of the amplitude-type FZP is relatively low ($\approx$ 9.7%). The focused intensity of the binary-phase FZP ($\approx$ 38.7%) increases by a factor of $\approx$ 4.0 times than that of the amplitude-type element. The intensity of the 4-level FZP (67.9%) is enhanced $\approx$ 7.0 times. The 8-level FZP structure can obtain 80.8% focusing efficiency, an increase factor of $\approx$ 8.3. The 16-level FZP has achieved the highest focusing efficiency at 91.6% of input energy (an increase factor $\approx$ 9.5). These values are all in close agreement to the theoretical calculations. The focusing efficiency of $\approx$ 91.6% obtained from the 16-level FZP is the highest value which has been obtained from these DOEs, which is suitable in practical applications.

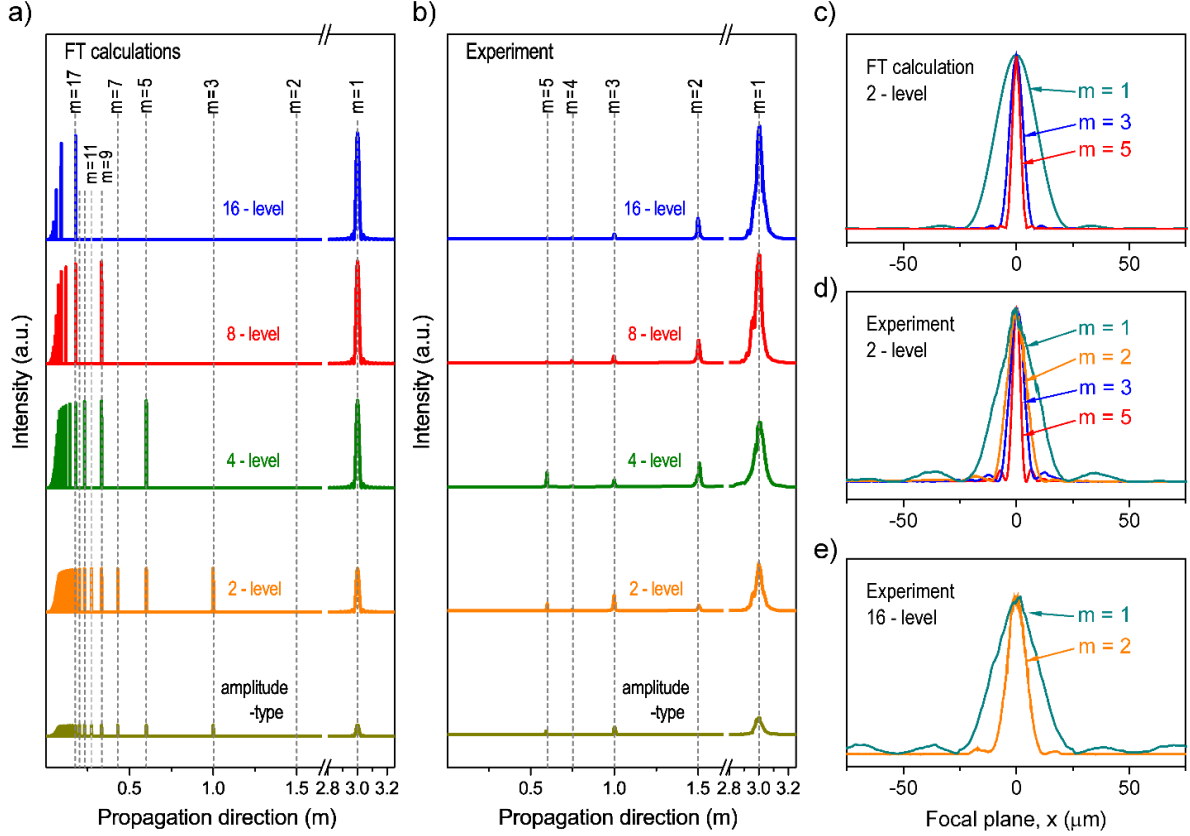

**Figure 6:** Intensity along the optical axis of MLFZPs. a) Theoretical intensity along the optical axis is computed using the Huygen – Fresnel approximation for DOEs. b) Experimental results of the intensity collected at different diffraction orders from DOEs including the amplitude-type and multilevel phase-type FZPs. c-e) Intensity profiles of different diffraction orders describe the focusing of light into focal planes, including theoretical calculations, 2- and 16-level FZPs, respectively.

In optical systems, the greatest challenge is to efficiently concentrate energy to obtain a small focal spot or a high resolution in point-to-point imaging.[1] The focal spot of a near plane wave incident on a focusing optical element is defined as the point spread function (PSF) and has a principal limit due to diffraction. For perfect optics governed by diffraction, the optical resolution or the spread of the diffraction-limited PSF of an imaging system is approximated by the first null of the Airy disk, corresponding to the Rayleigh resolution limit.[1] The focal spot size is limited only by the wavelength of the light ($\lambda$) and the numerical aperture (NA) or the F-number, $f_{number} = f/D$, in the form, $w_{null} \approx 1.22 \times \lambda \times f_{number} \approx 0.61 \times \lambda/NA$, for an optical system in air. We plot the intensity profile of our FFT calculations and experimental results as a function of the diffraction limit, $\lambda/NA$, in Figure 5 c,d. Under the 532-nm illumination (Figure 5c), a focal spot size with the full-width at half-maximum (FWHM), $w \approx (0.58 \pm 0.05) \times \lambda/NA$, has been obtained from polyimide membrane MLFZPs, indicating the diffraction limited performance of the DOEs. This value matches well with the FFT calculations of $w = 0.49 \times \lambda/NA$, demonstrating a remarkable fabrication accuracy of both the optical lithography on silicon molds and the polyimide casting processes. The minor discrepancy may originate in unavoidable fabrication errors and aberrations in our optical setup. Note that the diffraction limitation of our phase-type MLFZPs shows the same order when compared with those from reported multilevel phase supercritical lens of 0.41 $\lambda/NA$.[28] However, by employing the MLFZP pattern, the intensity at side-lobes is significantly suppressed and the energy is mainly concentrated into the primary focus, when compared with those from the supercritical lens design.[28]

Reducing the size of the focal spot for a specific wavelength of light can be achieved by increasing the numerical aperture, NA, or decreasing the F-number, $f_{number}$. Planar diffractive optics provide a great opportunity to increase NA values that are not easy to obtain in reflective and refractive systems by fabricating optical elements with a shorter focal length and/or large diameter. For optical characterization, a reduction in the $f_{number}$ ($f/D$) of a DOE can be realized by increasing the diameter of the incident beam on the DOE; thus, the resulting in narrowing of the focal spot size can be observed (Figure 4, inset). The black line is the focal spot size estimated from FFT calculations, while red symbols are experimental results. The focal spot size increases linearly with the $f_{number}$ of the DOEs, indicating a uniformity of the diffraction effect of our polyimide membrane MLFZPs from the center to outermost zones (the largest feature size is at the center and smallest structure is on the outermost ring). For a larger diameter of 0.080 m, a FWHM of the PSF of 23.8 ± 2.0 µm for the 16-level FZP with a focal length of 3 m was measured. This FWHM value is close to the theoretical value of 20.2 µm, demonstrating highly accurate fabrication processes.

Planar DOE's structures control optical power, and consequently alter the convergence of the incident beam and act as lenses to form images. To optimize the nanostructures of DOEs, and ultimately to efficiently collect light and form images, the optical properties of these features, including the energy distribution along the optical axis and the focusing efficiency at different diffraction orders need to be characterized. We have performed numerical calculations of the intensity profile along the propagation direction ($z$-axis) for the amplitude- and phase-type MLFZPs and compared them to our experimental results (Figure 6). The simplest diffractive element is the amplitude-type or conventional FZP. In this structure, the sequential transparent zone, $n^{th}$, is constructed in such a way that the addition to the optical path length to the primary focus is $n\lambda/2$, and thus, radiation adds in-phase to the first diffraction orders ($m = \pm 1$). Similar situations occur in higher diffraction orders ($m = \pm 2, \pm 3 \dots$) by adding optical path lengths of $mn\lambda/2$, which corresponds to focal lengths of $f_m = f/m$. Note that negative diffraction orders give rise to virtual foci with negative focal lengths. The diffraction efficiency for the $m^{th}$ diffraction order, $\eta_m = I_m/I_0$, is given by,[1, 27, 51, 56]

$$\eta_m = \begin{cases} 1/4 & m = 0 \\ 1/(m\pi)^2 & m \text{ odd} \\ 0 & m \text{ even} \end{cases}. \tag{9}$$

For a FZP with equal areas of transparent and opaque rings, half of the incident intensity is blocked by opaque areas. A fraction of 25% of the incident energy transmitted through the transparent areas goes to the zeroth order, producing a large background, and significantly reduces the signal-to-noise ratio at the focal plane. The sum of focused energy over all diffraction orders of an amplitude-type FZP is 25% of the input. Only a fraction ($\approx 10\%$) of the total incidence is diffracted to each of the first orders (10% goes to the first converging order, $m = 1$, another 10% goes to the diverging $m = -1$ order), a portion of $\approx 1\%$ goes to the 3$^{rd}$ order ($m = 3$), and so forth. Even diffraction orders will cancel each other at the focal position. However, even diffraction orders do appear for asymmetric widths of successive areas. For example, a zone plate where transparent alternate rings are narrower than the prescribed design due to imperfections in the fabrication process will show focal spots at even diffraction orders, depending on the degree of asymmetry (Figure 6).

In the case of a binary-phase diffractive lens (2-level FZP), the focusing efficiency increases significantly when opaque areas of the amplitude-type FZP become transparent and produce a phase shift of exactly π. In this structure, the zeroth order has no intensity. Due to the symmetrical geometry of the amplitude-type and binary-phase DOEs, light diffracts in a symmetrical way to all diffraction orders. Thus, for normal incidence the incident energy distributes equally to the positive and negative diffraction orders. The negative diffraction order, -$m$, produces a diverging wave, and the positive diffraction order, +$m$, generates a converging wave, both with equal focal lengths. As a result of this behavior, a binary-phase FZP lens acts at the same time as a diverging and converging lens, which is a drawback of binary-phase FZPs for imaging systems because the focused energy of the diverging diffraction order is equal to that of the converging order, and thus, collects non-useful light from objects, and remarkably reduces the signal-to-noise ratio at the plane of interest.[8] From our calculations, the maximum focusing efficiency of such a diffractive lens is 40.5% to each of the first orders (40.5% goes to the diverging diffraction order, $m = -1$, another 40.5% goes to the converging order, $m = 1$). The rest of the energy splits over higher diffraction orders. Note that the intensity at higher diffraction order modes of FFT calculation plots indicates the peak power (Figure 6a), not the total energy dropped in these modes. For these modes, the spot sizes at different diffraction orders need to be considered. The spot size at the $m^{th}$ diffraction order is reduced by a factor of $m$, causing an energy reduced by a factor of $m^2$ when compared with those of the first order. Focusing efficiencies for other diffraction orders are provided in Table S1 (Supporting Information). The measured focusing efficiency of our polyimide membrane binary-phase FZPs at the primary converging order ($m = 1$) is 38.7 ± 3.0%, which is very close to the theoretical value, demonstrating an accurate fabrication process.

A small signal from the 2$^{nd}$ diffraction order is detected from the binary-type FZPs at 1.5 m, likely due to unavoidable fabrication errors and inaccurate optical characterization. The intensity at the 3$^{rd}$ and 5$^{th}$ diffraction orders is higher than that at the 2$^{nd}$ order, which agrees with the pattern extracted from the FFT calculations.

Breaking the structural symmetry, such as multilevel phase-type structures of DOEs or a tilted incident beam, will produce different focusing efficiencies of the diverging and converging diffraction orders. It is important to point out that DOEs can be tailored to obtain a high focusing efficiency of a specific diffraction order, $m$, at a specific wavelength of light. Our calculations show that the focusing efficiency of the 4-level phase-type FZP can reach 81.1, 3.2 and 1.0% for the 1$^{st}$, 5$^{th}$, and 9$^{th}$ converging diffraction orders, respectively, with no intensity for the 3$^{rd}$, 7$^{th}$, 11$^{th}$, … and even orders (Figure 6a). For the diverging diffraction orders, the energy will be accumulated a portion of 9.1, 1.65, and 0.7% for the -3$^{rd}$, -7$^{th}$, -11$^{th}$, … diffraction orders, respectively. Specifically, we expect to observe light at ± 4$m$ + 1 diffraction orders with $m$ = 0, 1, 2..., in which the main input energy drops at the primary focus.

A high focusing efficiency can be obtained for higher-level FZP structures at the primary focus (Figure 6). For the 8-level FZP structure, the light can be collected at the 1$^{st}$ and ±8$m$+1 orders with $m$ = 1, 2,... Theoretically, the focusing efficiency can reach 95.0 and 1.1% for the 1$^{st}$ and 9$^{th}$ diffraction orders, respectively. Correspondingly, the energy drops at the -7$^{th}$, -15$^{th}$, … diverging diffraction orders by 1.9 and 0.4%, respectively. The focusing efficiency of the 16-level phase-type FZP can reach 98.7% at the primary converging diffraction order and almost zero for higher diffraction orders (±16$m$ + 1 orders with $m$ = 1, 2, ...). Thus, when the number of levels is 16 or higher, the phase-type MLFZPs can be considered as nearly continuous surface-relief elements. Indeed, it is not necessary to fabricate DOEs with more than 16 levels using photolithography techniques, because misalignment errors during fabrication in the lateral direction and inaccuracies in the etching depth will reduce the diffraction efficiency. For perfect fabrication, the increase in the focusing efficiency is very small (a fraction of a percent) when pushing forward the fabrication beyond the 16-level to a higher-level structure (32, 64 or 128-level FZPs); therefore, the added complexity of fabrication would not balance the small efficiency gain. Calculations for higher-level structures are provided in Figure S7, supporting information. We have obtained focusing efficiencies of 80.8 ± 3.0% and 91.6 ± 3.0% for the 8- and 16-level FZPs, respectively, at the primary focus, demonstrating an accurate nanofabrication process. These values are slightly lower than the theoretical values due to imperfections in the nanofabrication process, the flexibility of polyimide membranes, and imprecise optical characterization.

PSFs extracted from our calculations and experiments are critical for development, design, fabrication, and optical analysis of MLFZPs. The focal spot size or optical resolution, $w$, of a FZP is limited by the width $\Delta R_n$ of the outermost zone ($\approx 0.61\Delta R_n$). However, when working with the higher diffraction order, $m$, the focusing efficiency reduces with a factor of $m^2$ and the spot size decreases with a value of $\approx m$. Figures 6c,d, and e show normalized intensity profiles of FFT calculations and experimental results of polyimide membrane 2- and 16-level FZPs, respectively. As can be seen, the FWHM of the intensity profile at the $m^{th}$ diffraction order is narrower by the factor of $m$ for both FT calculations and experimental results. An excellent agreement between the measurements and calculations has been observed, which confirms the validity of the computational techniques. The spot sizes are narrower at high diffraction orders; thus, one might exploit these higher diffraction orders to achieve higher resolution imaging. In addition, calculations and experimental measurements of intensity profiles along the propagation direction (optical axis) of the 16-level MLFZP are added in the supporting information Figures S8 and S9.

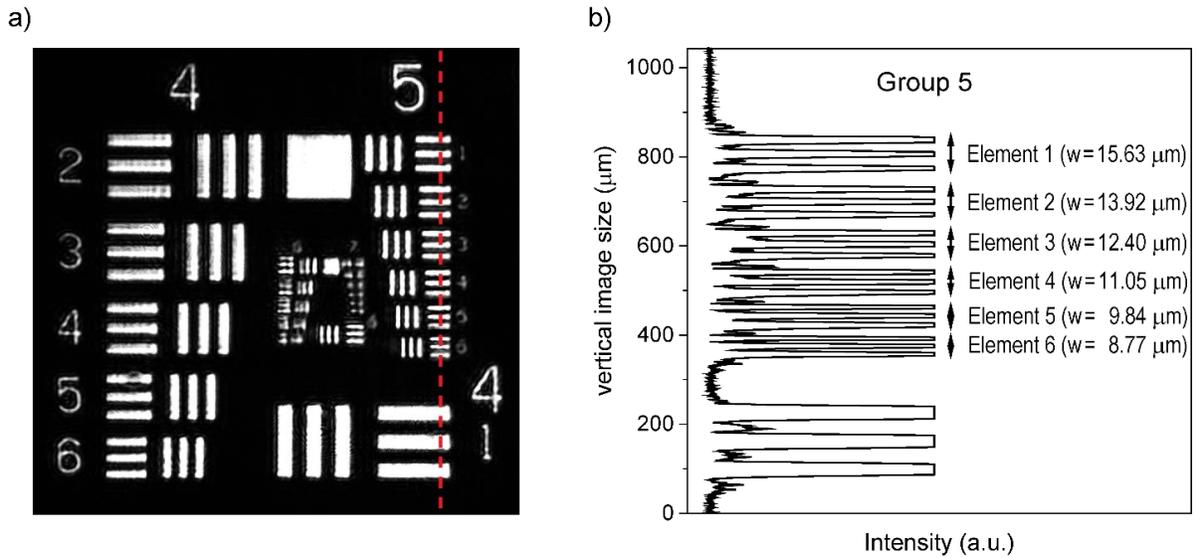

**Figure 7:** Imaging performance of the 16-level MLFZP. a) The resolution test target is imaged with the 16-level MLFZP at the wavelength of 532 nm, revealing features in groups 4 and 5 of the USAF 1951 test target. b) The intensity profile of the imaged target along the red dashed line in (a) shows resolved features in group 5.

To evaluate the imaging performance of the MLFZP, the collimation lens of the beam expander was replaced by an USAF 1951 glass slide test resolution target (clear pattern on a chrome background). The image of the resolution target was directly received by a CMOS camera (monochrome BFS-U3-200S6M-C with a pixel size of 2.4 × 2.4 µm) at the primary focus of the 16-level phase-type MLFZP (Figure 7). Elements in the group number 5 of 1, 2, 3, 4, 5, and 6 with the line width, $w$, of 15.63, 13.92, 112.40, 11.05, 9.84 and 8.77 µm, respectively, in the resolution target were observed clearly, as indicated by the red dashed line in Figure 7a with the intensity profile in Figure 7b. Element 6 in group 5 which contains 57-line pairs per mm with a width of 8.77 µm was resolved as shown by the intensity profile in Figure 7b, indicating that the resolution of the imaging performance is better than 8.77 μm with 532-nm light. The result is consistent with diffraction-limited performance of 23.8-μm resolution as shown in Figure 5. The imaging results demonstrate that the fabricated MLFZP shows potential as a planar focusing optic that can be fabricated on flexible materials and integrated in large aperture imaging systems.

## 5. Conclusion

In summary, we have successfully fabricated and characterized phase-type multilevel DOEs by employing nanofabrication processes including photolithography and polyimide membrane casting. The MLFZPs have a significantly enhanced focusing efficiency, as predicted, with an efficiency of $91.6 \pm 3.0\%$ for the 16-level DOE. The focusing efficiency of these lenses is higher than any previous report on flexible materials, and their optical performance agrees well with simulations. Additionally, large aperture DOEs can be fabricated using this technique by arranging them into a hexagon configuration. These MLFZPs exhibit a weak intensity at their side-lobes and considerably reduced unwanted diffraction orders, thus yielding an extremely low background signal. Moreover, these lenses have been fabricated at a significantly reduced time and cost relative to other techniques. The results suggest that polyimide membrane MLFZPs are appropriate for various sensing and imaging applications, where previous efficiency limitations prevented their application.

**ACKNOWLEDGMENTS.** The authors gratefully acknowledge the financial support of this effort by the Earth Science Technology Office (ESTO), NASA Science Mission Directorate, and NASA Langley Research Center.

**Supporting Information.**

Supporting Information is available from the Wiley Online Library or from the author.

**Conflict of Interest**

The authors declare no conflict of interest.

**Data Availability Statement**

The data that support the findings of this study are available from the corresponding author upon reasonable request.

# Supporting Information

# High-Efficiency Multilevel Phase Lenses with Nanostructures on Polyimide Membrane

Leslie Howe,[1] Tharindu D. Rajapaksha,[1] Kalani H. Ellepola,[1] Vinh X. Ho,[1] Zachary Aycock,[1] Minh L. P. Nguyen,[1] John P. Leckey,[2] Dave G. Macdonnell,[2] Hyun Jung Kim,[2] and Nguyen Q. Vinh[1]*

[1] Department of Physics and Center for Soft Matter and Biological Physics, Virginia Tech, Blacksburg, VA 24061, USA

[2] NASA Langley Research Center, Hampton, Virginia 23681, USA

* Corresponding author: vinh@vt.edu; phone: 1-540-231-3158

## 1. Fabrication processes

The fabrication of nanostructures for diffractive optics has been accomplished through a variety of differing techniques, such as laser ablation,[1] electron beam (e-beam),[2] and photolithography.[3] In this work, we have employed a cooperative nanofabrication technique based on the photolithography and polyimide membrane casting processes for realizing amplitude-type and phase-type multilevel Fresnel zone plates (MLFZPs).[4, 5] To fabricate the amplitude-type FZP, we use standard photolithography (Figure S1). First, a quartz wafer is coated with photoresist by spin coating. The photoresist is exposed to ultraviolet light with a pattern from a photomask. The exposure to ultraviolet light causes a chemical change in the photoresist that can be removed by a special solution or developer. Then, the wafer with the photoresist pattern is covered with a 75-nm chromium layer using e-beam evaporation deposition (Kurt J Lesker PVD-250). Finally, the wafer is submerged in acetone for the lift-off process. [5, 6]

For the phase-type MLFZP fabrication, two cooperative processes have been applied to obtain desirable nanostructures on polyimide membranes including the photolithography and membrane casting processes. The MLFZPs have been fabricated by casting CORIN polyimide membrane on silicon molds. To begin the fabrication process of silicon molds, we bond hexamethyldisilane (HMDS) to the surface of a silicon wafer through spin coating and soft bake at 110°C for 4 minutes. Through this heating process, HMDS chemically bonds to an oxidized surface to allow for the photoresist to adhere. We used a high-resolution photoresist that is optimal for high resolution and high thermal stability for chemical etching. By depositing a micrometer thick layer and soft baking at 90°C for 1 minute, we can expect the feature resolution from this photoresist of around 0.6 $\mu$m. Following the deposition of photoresist by spin coating, we expose the wafer using a Karl Suss MA6 with an *i*-line and a dosage of 24 – 30 mJ/cm$^2$, depending on the minimum feature size we want to obtain. For larger features, a higher dosage allows for more distinct edges on the pattern, while to ensure the correct resolution for small features, a smaller dose is necessary. The alignment is done using multiple markers along the edges of the features to allow for the highest precision, along with a hard contact exposure and minimum allowable alignment gap between the mask and substrate (Stern). To fabricate silicon molds for MLFZPs, we use several photomasks. The first photomask shapes 2-level zones, and each consecutive mask will divide the zones in half for further levels, until 16-level structures are achieved. With this photolithography method, we can develop features down to the ~1 µm level. After developing the photoresist, the silicon wafer is etched down to the expected depth. This step is done using reactive ion etching (RIE), which is a multi-step plasma-induced process[7] First, process gases are introduced into the etch chamber. In this case, we use $SF_6/O_2/CHF_3$ with flow rates of 20/10/10 sccm, respectively. After these gases are broken down into chemically reactive particles by the plasma, these radicals diffuse to the surface of the substrate and are absorbed by the silicon material layer. This causes a reaction between the radicals and silicon, which will result in reaction byproducts being diffused and cycled out of the chamber. Our procedure uses a pressure of 20 mTorr and a substrate temperature of 5°C, with a high power of 100 W, resulting in an etch rate of about 4 nm/s. For the 2-level FZP, we use an etch depth of 490 nm. The 4-level uses a step size of 245 nm, the 8-level has a step size of 123 nm, and the 16-level undergoes a step size of 62 nm. After every etching step, we return to photolithography. For example, after the etching step for the 2-level, we use photolithography, fabricate the 4-level step, and continue the process until complete. These structures have been evaluated with a JEOL IT-500HR scanning electron microscope (SEM) to make sure that the error is within ± 5 nm. Finally, the CORIN polyimide membrane is cast into the silicon molds. After several steps including peeling off the polyimide membrane from the silicon molds, and post-treatment, the MLFZP lenses are fully functional and their optical properties were characterized in our optical setup.

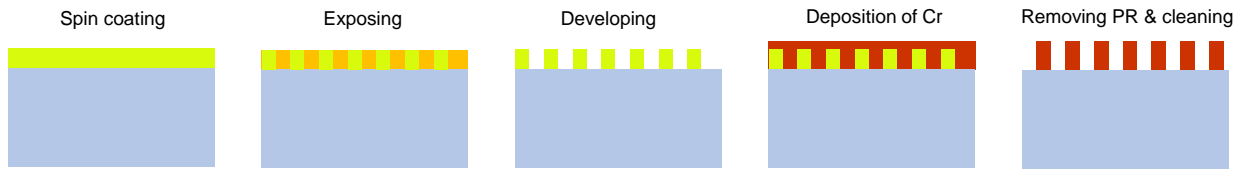

**Figure S1.** Photolithography steps to fabricate amplitude-type FZP on a quartz wafer.

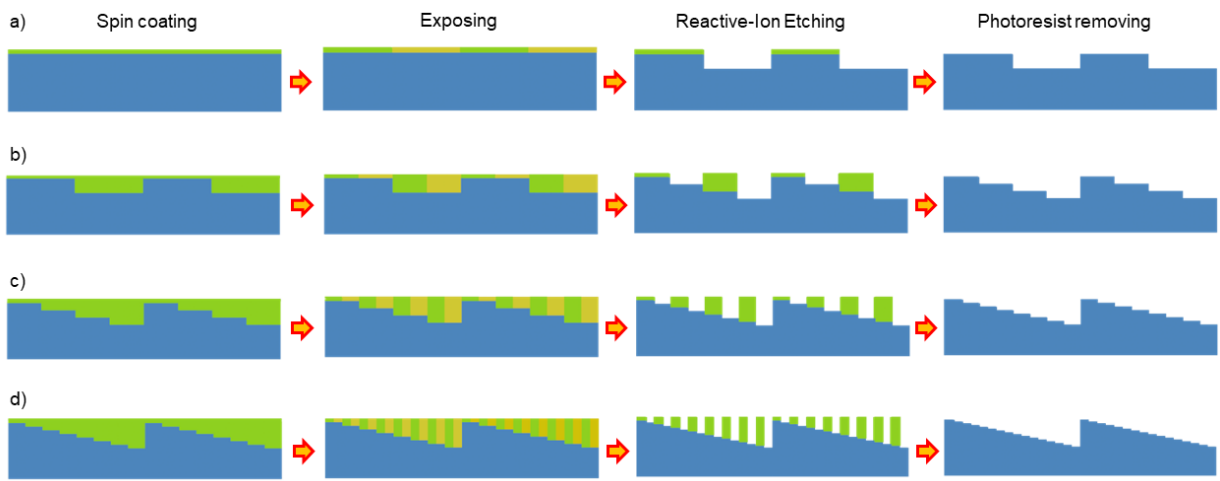

**Figure S2.** Photolithography steps for fabricating silicon molds to produce 2, 4-, 8-, and 16-level FZPs.

## 2. Optical and scanning electron microscope images

To obtain highly accurate nanofabrication, it is crucial to view these dimensions holistically in both the profile and depth of silicon molds, involving several characterization methods including optical imaging, electron imaging, and contact profilometry. The physical profile, obtained using atomic force microscope (AFM) images (Bruker DektakXT and Bruker Icon AFM), can be accurately measured to a theoretical limit of 4 Å, depending on setting parameters used and system variation. With this, we are able to accurately and repeatedly report the smallest step size of around 50 nm, and therefore determine the etching rate. While this method of contact profilometry is dependable and informative for vertical dimensions, another dimension of importance is the anisotropy of the etched edge, in which horizontal measurement is less precise using contact methods. Obtaining a 90° angle is important to effectively diffract the light, so in our efforts to find a recipe which could achieve this, we need to measure the sidewall angle. To do this, we have employed scanning electron microscopy (SEM). The SEM allows us to image the smallest features clearly and accurately, so we may understand how the sidewall is affected by certain etching parameters. Combined with profilometry, this measurement gives us an accurate representation of both the depth and the character of the etch.

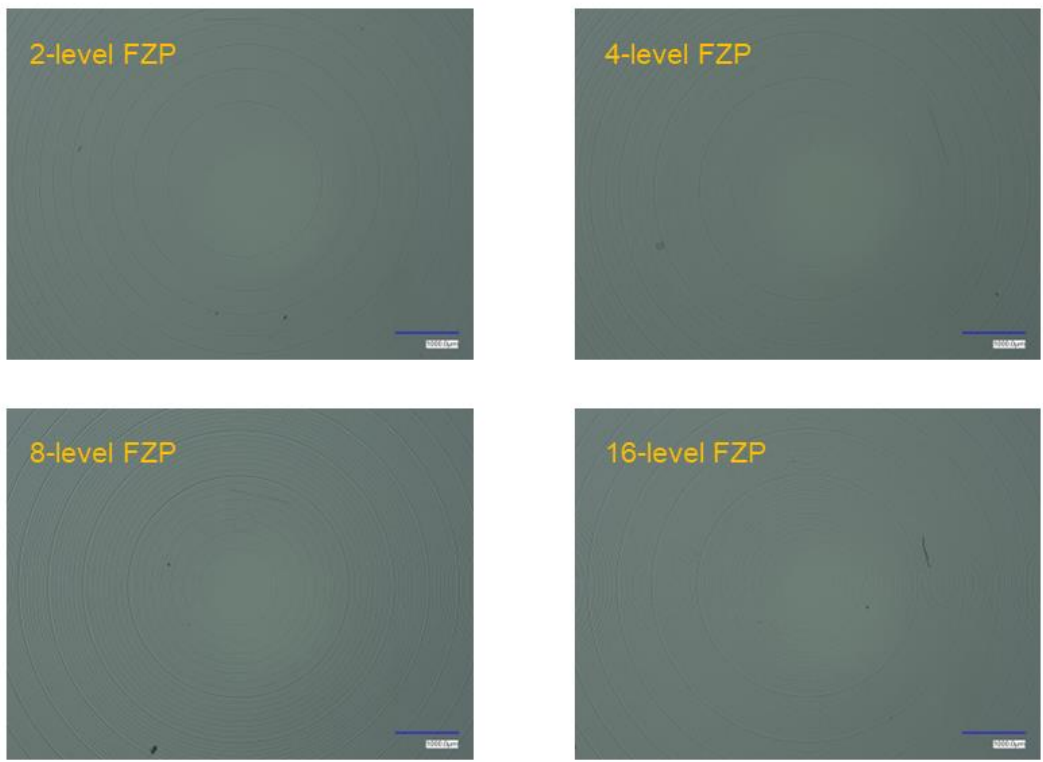

**Figure S3.** Optical images of silicon molds for fabricating 2, 4-, 8-, and 16-level FZPs.

The final dimension that is crucial to high performance lenses is the zone width. Because this may be affected by both the methods used in photolithography as well as plasma etching, this variable must be strictly controlled. Optical imaging has been employed to inspect the quality and size of the resist. This is done using a high efficiency optical microscope (Keyence VHX-700), which allows for measurements of feature sizes with a magnification of up to 6000×. These methods of physical characterization allow us to remain within the necessary parameters and create highly efficient imaging lenses. Figure S3 shows optical microscope images of silicon molds for fabricating phase-type MLFZP (2, 4-, 8-, and 16-levels). Figure S4 provides SEM images of silicon molds for fabrication of 2- and 16-level FZLs. Figure S6 shows optical microscope images of phase-type MLFZP (2, 4-, 8-, and 16-levels).

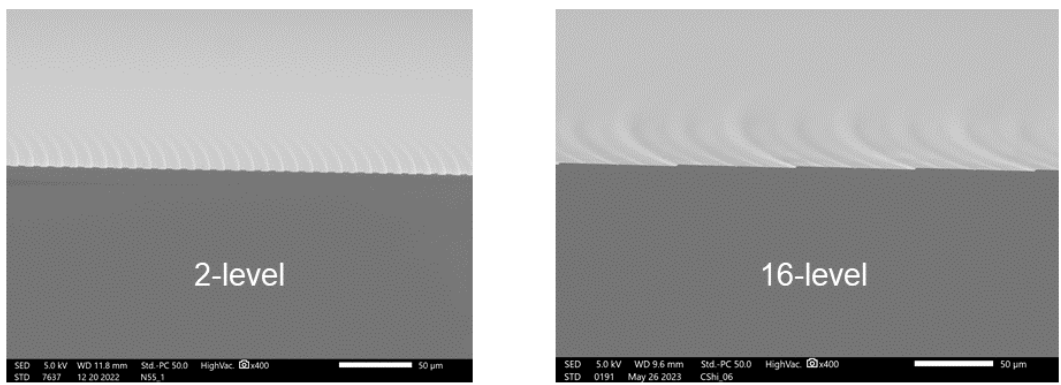

**Figure S4.** SEM images of silicon molds for fabricating 2- and 16-level FZPs.

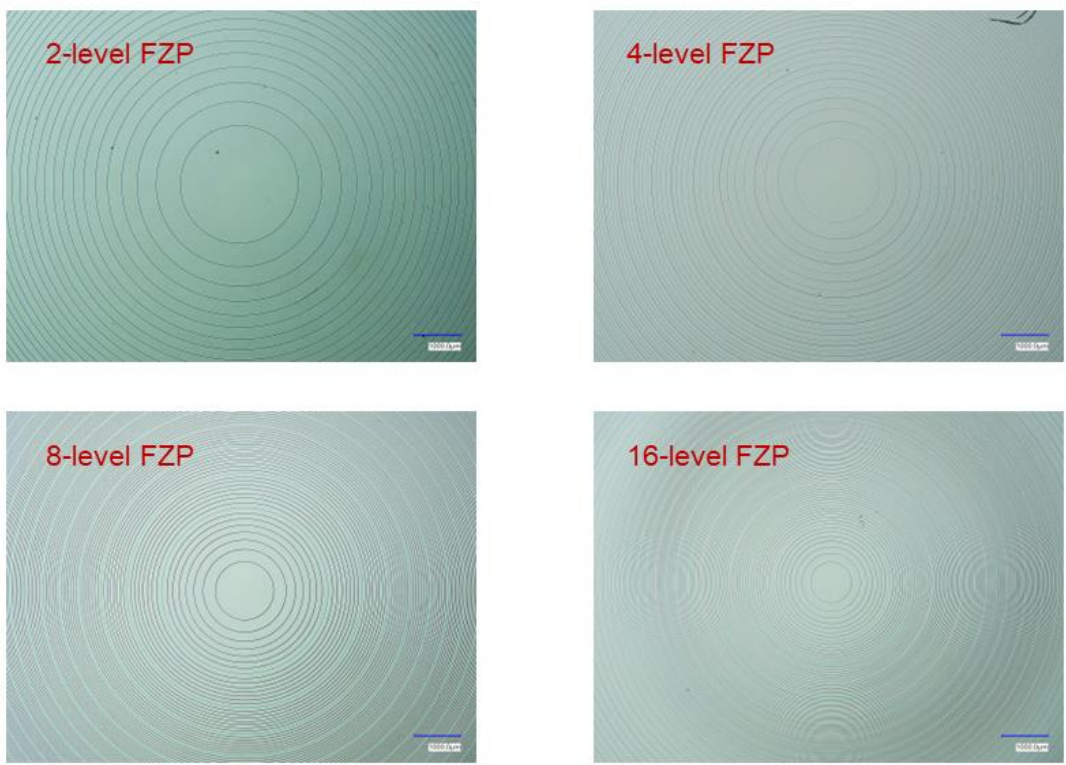

**Figure S5.** Optical images of plastic membrane 2, 4-, 8-, and 16-level FZPs.

## 3. AFM images and profiles

We characterize the height of the nanostructure on silicon molds as well as polyimide membrane MLFZPs using AFM. Figure S6 shows AFM images and profiles of silicon molds for fabricating 2, 4-, 8- and 16-level FZPs.

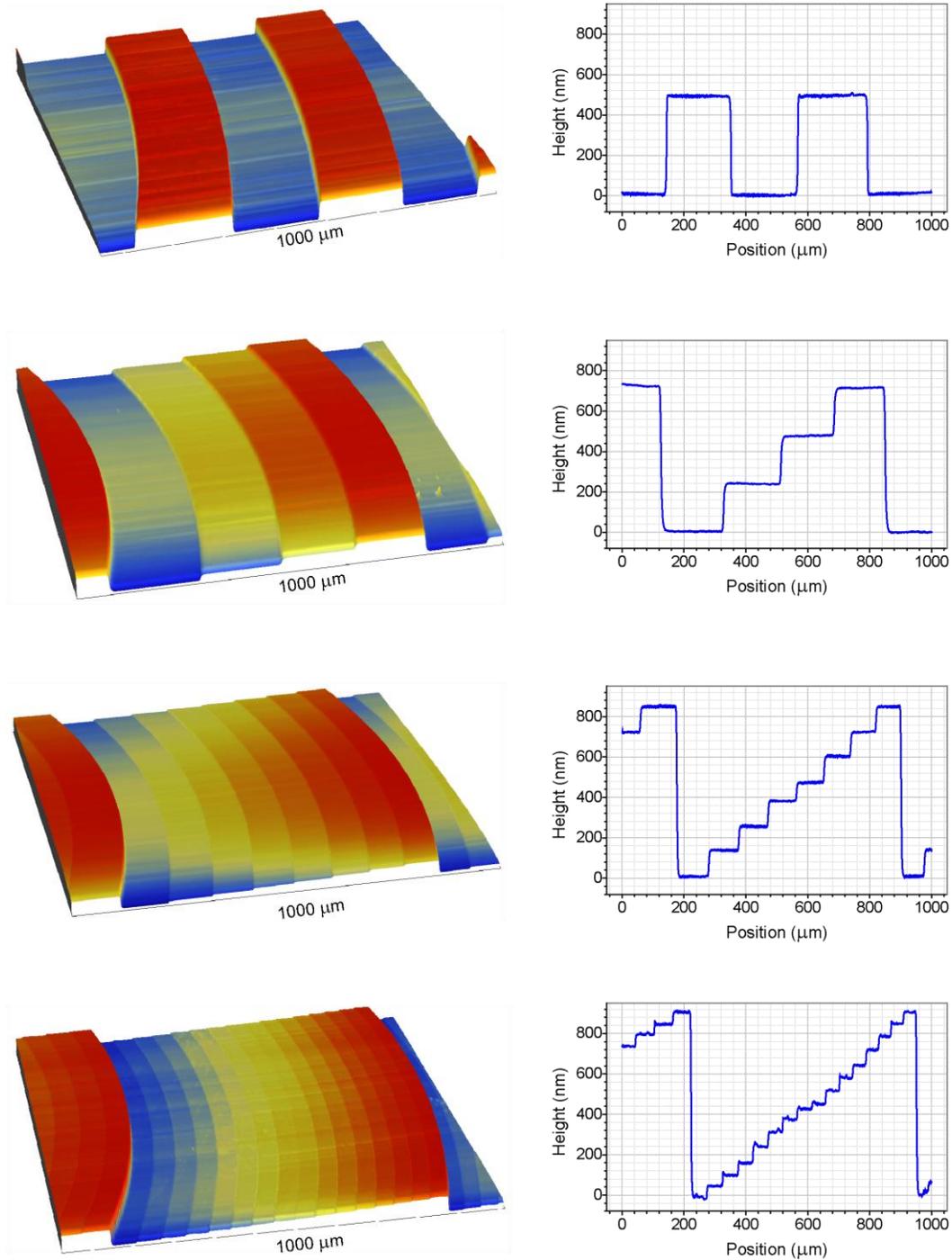

**Figure S6.** AFM images and vertical profiles of silicon molds at a position close to the center.

## 4. Focusing efficiency at different diffraction orders

We have calculated the focusing efficiency at different diffraction orders using FFT in MATLAB. Table S1 provides focusing efficiencies for amplitude-type, phase-type MLFZPs (2-, 4-, 8- and 16-levels) at different diffraction orders.

**Table S1:** Focusing efficiency at different diffraction orders (values in percentage).

| Diffraction orders ($m$) | Amplitude-type | | Phase-type MLFZPs | | | | | | | |
| --- | --- | --- | --- | --- | --- | --- | --- | --- | --- | --- |
| | | | 2-level | | 4-level | | 8-level | | 16-level | |
| | positive | negative | positive | negative | positive | negative | positive | negative | positive | negative |
| 1 | **10.13** | **10.13** | **40.53** | **40.53** | **81.06** | 0.00 | **94.96** | 0.00 | **98.72** | 0.00 |
| 2 | 0.00 | 0.00 | 0.00 | 0.00 | 0.00 | 0.00 | 0.00 | 0.00 | 0.00 | 0.00 |
| 3 | **1.13** | **1.13** | **4.50** | **4.50** | 0.00 | **9.01** | 0.00 | 0.00 | 0.00 | 0.00 |
| 4 | 0.00 | 0.00 | 0.00 | 0.00 | 0.00 | 0.00 | 0.00 | 0.00 | 0.00 | 0.00 |
| 5 | **0.41** | **0.41** | **1.62** | **1.62** | **3.24** | 0.00 | 0.00 | 0.00 | 0.00 | 0.00 |
| 6 | 0.00 | 0.00 | 0.00 | 0.00 | 0.00 | 0.00 | 0.00 | 0.00 | 0.00 | 0.00 |
| 7 | **0.21** | **0.21** | **0.83** | **0.83** | 0.00 | **1.65** | 0.00 | **1.94** | 0.00 | 0.00 |
| 8 | 0.00 | 0.00 | 0.00 | 0.00 | 0.00 | 0.00 | 0.00 | 0.00 | 0.00 | 0.00 |
| 9 | **0.13** | **0.13** | **0.50** | **0.50** | **1.00** | 0.00 | **1.17** | 0.00 | 0.00 | 0.00 |
| 10 | 0.00 | 0.00 | 0.00 | 0.00 | 0.00 | 0.00 | 0.00 | 0.00 | 0.00 | 0.00 |
| 11 | **0.08** | **0.08** | **0.33** | **0.33** | 0.00 | **0.67** | 0.00 | 0.00 | 0.00 | 0.00 |
| 12 | 0.00 | 0.00 | 0.00 | 0.00 | 0.00 | 0.00 | 0.00 | 0.00 | 0.00 | 0.00 |
| 13 | **0.06** | **0.06** | **0.24** | **0.24** | **0.48** | 0.00 | 0.00 | 0.00 | 0.00 | 0.00 |
| 14 | 0.00 | 0.00 | 0.00 | 0.00 | 0.00 | 0.00 | 0.00 | 0.00 | 0.00 | 0.00 |
| 15 | **0.05** | **0.05** | **0.18** | **0.18** | 0.00 | **0.36** | 0.00 | **0.42** | 0.00 | **0.44** |
| 16 | 0.00 | 0.00 | 0.00 | 0.00 | 0.00 | 0.00 | 0.00 | 0.00 | 0.00 | 0.00 |
| 17 | **0.04** | **0.04** | **0.14** | **0.14** | **0.28** | 0.00 | **0.33** | 0.00 | **0.34** | 0.00 |
| 18 | 0.00 | 0.00 | 0.00 | 0.00 | 0.00 | 0.00 | 0.00 | 0.00 | 0.00 | 0.00 |
| 19 | **0.03** | **0.03** | **0.11** | **0.11** | 0.00 | **0.22** | 0.00 | 0.00 | 0.00 | 0.00 |
| 20 | 0.00 | 0.00 | 0.00 | 0.00 | 0.00 | 0.00 | 0.00 | 0.00 | 0.00 | 0.00 |
| 21 | **0.02** | **0.02** | **0.09** | **0.09** | **0.18** | 0.00 | 0.00 | 0.00 | 0.00 | 0.00 |
| 22 | 0.00 | 0.00 | 0.00 | 0.00 | 0.00 | 0.00 | 0.00 | 0.00 | 0.00 | 0.00 |
| 23 | **0.02** | **0.02** | **0.08** | **0.08** | 0.00 | **0.15** | 0.00 | **0.18** | 0.00 | 0.00 |
| 24 | 0.00 | 0.00 | 0.00 | 0.00 | 0.00 | 0.00 | 0.00 | 0.00 | 0.00 | 0.00 |
| 25 | **0.02** | **0.02** | **0.06** | **0.06** | **0.13** | 0.00 | **0.15** | 0.00 | 0.00 | 0.00 |
| 26 | 0.00 | 0.00 | 0.00 | 0.00 | 0.00 | 0.00 | 0.00 | 0.00 | 0.00 | 0.00 |
| 27 | **0.01** | **0.01** | **0.06** | **0.06** | 0.00 | **0.11** | 0.00 | 0.00 | 0.00 | 0.00 |
| 28 | 0.00 | 0.00 | 0.00 | 0.00 | 0.00 | 0.00 | 0.00 | 0.00 | 0.00 | 0.00 |
| Sum | 12.32 | 12.32 | 49.28 | 49.28 | 86.37 | 12.18 | 96.62 | 2.54 | 99.06 | 0.44 |
| Blocked | 50.00 | | | | | | | | | |
| Zero-order | 25.00 | | | | | | | | | |
| Total | | 99.64 | | 98.55 | | 98.55 | | 99.16 | | 99.50 |

## 5. Focusing efficiency of MLFZPs

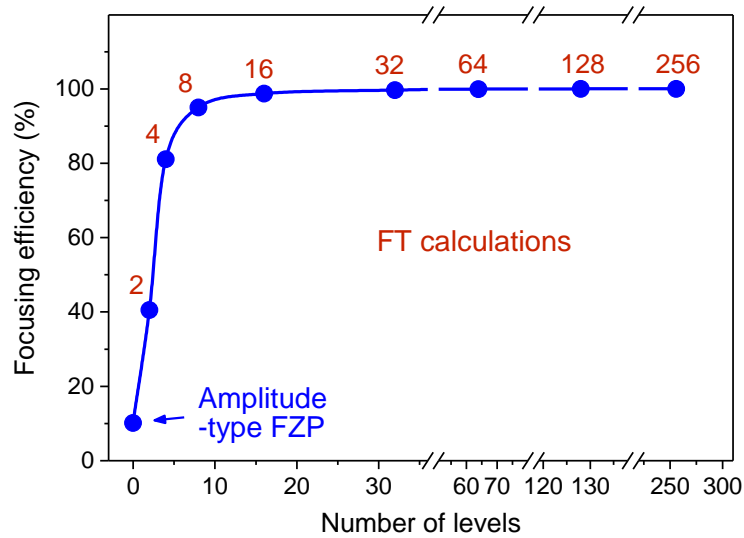

**Figure S7.** Theoretical calculations for intensity along the optical axis are computed using the Huygen – Fresnel approximation for DOEs, including the amplitude-type and multilevel phase-type FZPs from 2-level to 256-level.

## 6. Intensity profile along the propagation direct (optical axis) of the 16-level MLFZP

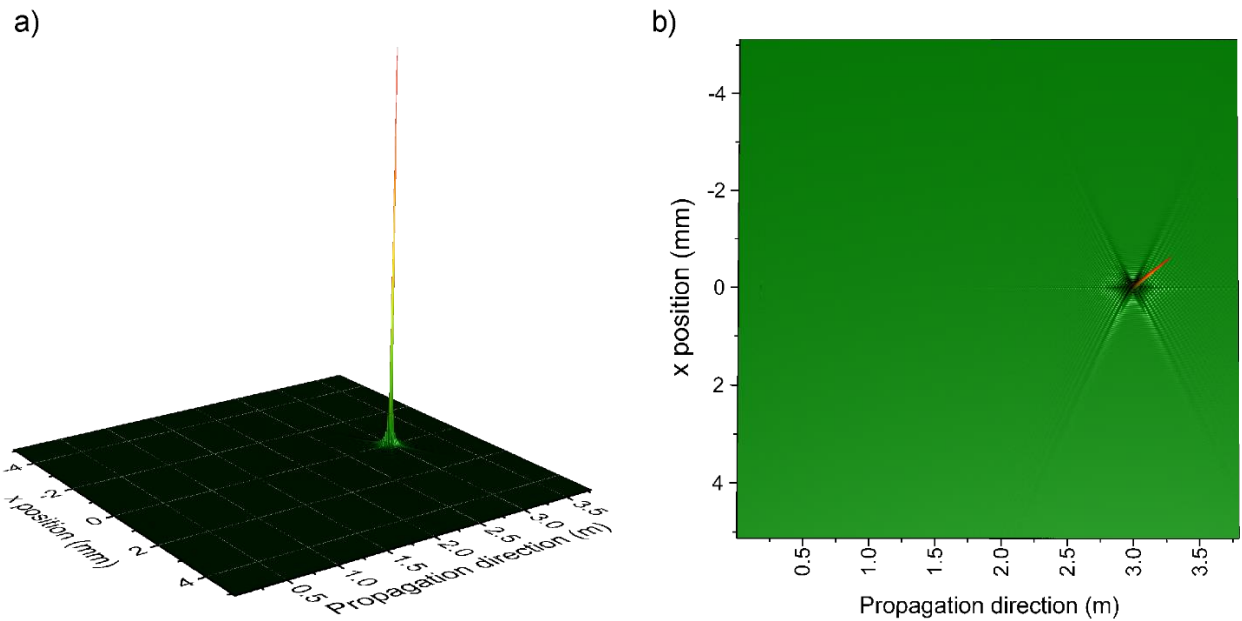

**Figure S8.** Fourier transform calculations for the intensity profile along the propagation direction (optical axis) using the Huygen – Fresnel approximation for the 16-level MLFZP (a) 3D color map surface and (b) flattened view of the intensity.

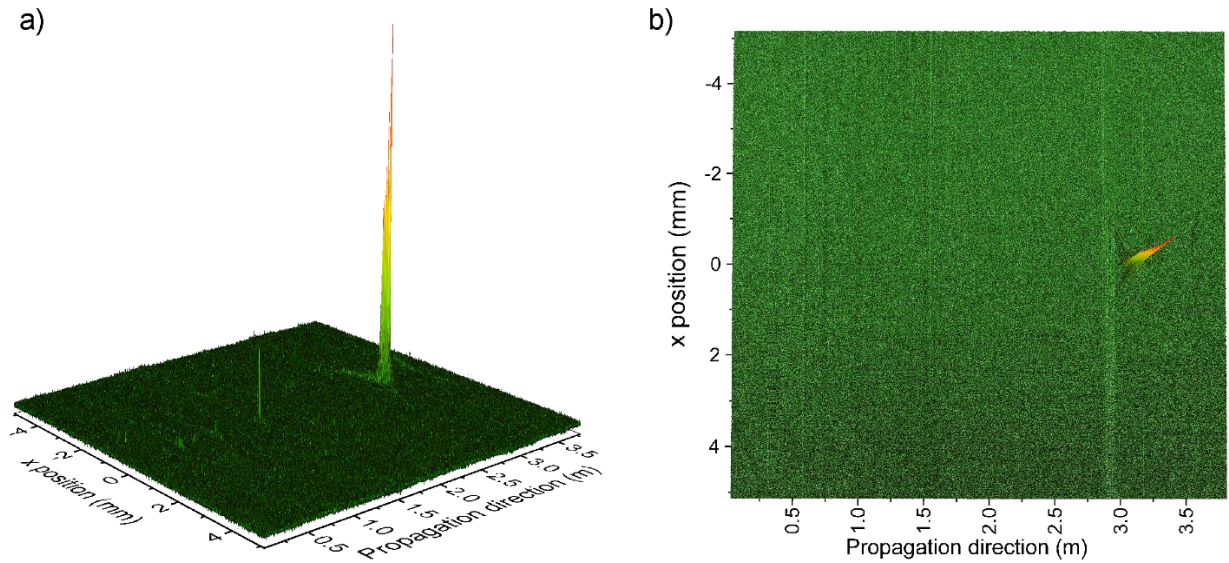

**Figure S9.** Experimental results of the intensity profile along the propagation direction (optical axis) of the 16-level MLFZP (a) 3D color map surface and (b) flattened view of the intensity.